\documentstyle[prl,aps,twocolumn]{revtex}
\input epsf
\begin{document}
\twocolumn[\hsize\textwidth\columnwidth\hsize\csname
@twocolumnfalse\endcsname

\draft

\title{Finite size errors in quantum many--body simulations of extended
systems}

\author{P.~R.~C.~Kent$^1$, Randolph~Q.~Hood$^1$,
A.~J.~Williamson$^1$\cite{address}, R.~J.~Needs$^1$,
W.~M.~C.~Foulkes$^2$, and G.~Rajagopal$^1$}

\address{$^1$Cavendish Laboratory, Madingley Road, Cambridge CB3 0HE,
UK}

\address{$^2$The Blackett Laboratory, Imperial College, Prince Consort
Road, London SW7 2BZ, UK}

\date{\today}

\maketitle

\begin{abstract}
\begin{quote}
\parbox{16 cm}{\small 
  Further developments are introduced in the theory of finite size
  errors in quantum many--body simulations of extended systems using
  periodic boundary conditions.  We show that our recently introduced
  Model Periodic Coulomb interaction [A.~J.~Williamson {\it et al.},
  Phys. Rev.  B {\bf 55}, R4851 (1997)] can be applied consistently to
  all Coulomb interactions in the system.  The Model Periodic Coulomb
  interaction greatly reduces the finite size errors in quantum
  many--body simulations.  We illustrate the practical application of
  our techniques with Hartree--Fock and variational and diffusion
  quantum Monte Carlo calculations for ground and excited state
  calculations.  We demonstrate that the finite size effects in
  electron promotion and electron addition/subtraction excitation
  energy calculations are very similar.}
\end{quote}
\end{abstract}


\narrowtext
]

\section{Introduction}

Most simulations of extended quantum systems are performed using
finite simulation cells.  This introduces ``finite size errors'' which
are one of the major problems limiting the application of accurate
many--body techniques to extended systems.  The standard method of
reducing the finite size errors is to apply periodic boundary
conditions, but important finite size errors often remain.  In this
paper we present further developments of the theory of finite size
effects in quantum many--body simulations subject to periodic boundary
conditions.  Our motivation is to understand and reduce the finite
size effects encountered in quantum Monte Carlo simulations, although
the techniques described here are of wide generality and can be
readily applied to other many--body electronic structure methods.

Quantum Monte Carlo (QMC) methods in the variational~\cite{vmc} and
diffusion~\cite{dmc,hammond} forms are capable of yielding highly
accurate results for correlated electron systems.  These methods are
very promising because (i) electron correlations are included
essentially without approximation and (ii) the methods scale
favourably with system size.  Nevertheless, for realistic systems the
cost of these calculations remains large, and it would be highly
desirable to reduce the finite size errors so that accurate results
can be obtained using small simulation cells.  We stress that all
quantum many--body calculations which use periodic boundary conditions
to model extended systems suffer from finite size effects and that the
ideas discussed in this paper are relevant whenever long ranged
interactions are involved.

Finite size errors in many--body calculations have traditionally been
corrected for by extrapolation techniques and/or by using the results
of more approximate calculations, such as those based on density
functional theory (DFT).  While extrapolation techniques can certainly
be successful they are very costly.  In addition, the finite size
errors in periodic boundary condition DFT calculations are
significantly different from those in the standard formulations of
quantum many--body techniques such as QMC, and therefore a simple
application of a DFT finite size correction may not lead to accurate
results.  We have understood the reason for this difference and have
found a way to reduce the finite size errors in quantum many--body
simulations using a new ``Model Periodic Coulomb'' (MPC) interaction,
which has the additional advantage that the residual finite size
effects are reasonably well described by standard DFT calculations.
The accuracy can be further increased by using an extrapolation
procedure, but the extrapolation corrections are considerably reduced
and can therefore be evaluated using a smaller range of system sizes.

The layout of this paper is as follows.  In section~\ref{sec:hamil} we
describe the Hamiltonian within periodic boundary conditions, while in
section~\ref{sec:fscorr} we discuss various finite size correction and
extrapolation procedures.  In section~\ref{sec:ipsereduction} we
review the ``independent particle finite size effects'', showing how
our k--space sampling techniques are related to those used in
mean--field theories.  In section~\ref{sec:cfse} we introduce our MPC
interaction for reducing finite size effects in periodic systems and
show that it can be applied to all the Coulomb interactions.  We
present tests of the MPC interaction within HF theory
(section~\ref{sec:hfapp}), VMC (section~\ref{sec:vmcapp}), and DMC
(section~\ref{sec:dmcapp}).  The latter section includes DMC results
for a system with 1000 electrons, which is the largest number in any
DMC calculation to date.  In section~\ref{sec:exen} we discuss finite
size errors present in calculations of excitation energies.  Tests
within HF theory are presented in section~\ref{HFexcite}, while in
section~\ref{QMCexcite} we present VMC results for the ``optical
absorption'' and ``photoemission'' gaps, the latter being the first
such calculations for a three--dimensional periodic system.
Calculations with up to 512 electrons demonstrate that the finite size
errors in the ``optical absorption'' and ``photoemission'' gaps are
similar and that the finite size effects are quite small even for a
64--electron simulation cell. We draw our conclusions in
section~\ref{sec:conclusions}.

\section{The Hamiltonian within periodic boundary conditions}
\label{sec:hamil}

The many--body Hamiltonian for a system of electrons at positions
${\bf r}_i$ and static ions at positions ${\bf x}_\alpha$
is~\cite{units}

\begin{eqnarray}\label{hamiltonian} \hat{H} & = &
\sum_{i}-\frac{1}{2}\nabla^2_i + \sum_i\sum_{\alpha}v_{\alpha}({\bf
r}_i,{\bf x}_\alpha) \nonumber \\ & + & \frac{1}{2}\sum_i\sum_{j\neq
i}v({\bf r}_i,{\bf r}_j) + \frac{1}{2}\sum_{\alpha}\sum_{\beta\neq
\alpha}v_{\alpha\beta}({\bf x}_{\alpha},{\bf x}_{\beta})\;\;.
\end{eqnarray}

\noindent An infinite system is normally modeled by a finite
simulation cell subject to periodic boundary conditions.  The model
interaction terms, $v_{\alpha}$, $v$, and $v_{\alpha\beta}$, are
chosen such that the potential energy of the model system, which
involves only the positions of the particles in the finite simulation
cell, mimics the potential energy of the infinite system as closely as
possible.  Since the potential energy of the infinite system depends
on the positions of all the charges in the solid, only a few of which
are included in the simulation, the model interaction energy is
approximate even in crystalline solids.  To enforce the periodic
boundary conditions the functions $v_{\alpha}$, $v$, and
$v_{\alpha\beta}$ must be invariant under the translation of either
argument by a member of the set of translation vectors of the
simulation cell lattice, $\{{\bf R}\}$. The standard approach is to
choose the model Hamiltonian such that the full potential energy of
Eq.~\ref{hamiltonian}, evaluated by summing the model interactions
between all pairs of particles in the simulation cell, equals the
potential energy per cell of an infinite array of identical copies of
the simulation cell.  However, even when we restrict each simulation
cell in the array of copies to be overall charge neutral, the sum of
inter--particle Coulomb $1/r$ interactions is only conditionally
convergent,~\cite{deleeuw} and to define this model interaction
uniquely the boundary conditions at infinity must be specified.  The
standard procedure is to define the potential by solving Poisson's
equation subject to periodic boundary conditions, in which case the
model interaction is the Ewald interaction.~\cite{Ewald} For two
electrons separated by ${\bf r}$, the Ewald interaction is

\begin{eqnarray}\label{ewald}
v_{{\rm E}}({\bf r}) & = & \frac{1}{\Omega} \sum_{\{{\bf G}\} \neq
{\bf 0}} \frac{4\pi}{G^2} \,{\rm exp} \left[-G^2/4\kappa^2 + i {\bf
G.r} \right] \nonumber \\ & - & \frac{\pi}{\kappa^2 \Omega} +
\sum_{\{{\bf R}\}} \frac{{\rm erfc}\left( \kappa | {\bf r} + {\bf R}|
\right)} {|{\bf r} + {\bf R}|} \;\;\;,
\end{eqnarray}

\noindent where $\Omega$ is the volume of the simulation cell, $\{{\bf
G}\}$ is the set of reciprocal space translation vectors of the
simulation cell lattice, and $\kappa$ is a positive but otherwise
arbitrary constant.

\section{Finite size correction and extrapolation formulae}
\label{sec:fscorr}

The basic idea behind finite size correction formulae is to write the
energy for the infinite system as

\begin{equation}
\label{fscorrection}
E_{\infty} = E_{N} + (E_{\infty} - E_{N})\;,
\end{equation}

\noindent where the subscript denotes the system size.  A highly
accurate many--body calculation for the $N$--particle system is then
performed to obtain an approximation for $E_{N}$, and the correction
term in brackets is approximated using a much less expensive scheme
which can be applied to very large systems.

In an extrapolation procedure the energy $E_{N}$ is calculated for a
range of system sizes and fitted to a chosen functional form which
contains parameters.  Correction and extrapolation procedures can be
combined to give an expression

\begin{equation}
\label{fs_corr-extrap}
E_{\infty} \simeq E_{N} + (E^{\prime}_{\infty} - E^{\prime}_{N}) + F(N)\;,
\end{equation}

\noindent where the prime indicates that a less expensive scheme is
used and $F(N)$ is an extrapolation function. Clearly, the optimal
form of $F(N)$ depends on the method used to calculate the correction
term.  The practical difference between correction and extrapolation
is that correction requires a single calculation of $E_{N}$ using the
accurate (and normally very costly) many--body technique, while
extrapolation requires several such calculations for different values
of $N$ and a subsequent fit to the chosen functional form $F(N)$. The
extrapolation procedure is costly because it involves more
calculations and is prone to inaccuracy because one has to perform a
fit with only a few data points. In designing a
correction/extrapolation procedure one therefore tries to make the
extrapolation term as small as possible.

Candidate methods for evaluating the correction term include
Kohn--Sham DFT and HF theory.  The most convenient methods are the
local density approximation (LDA) to DFT, or extensions such as the
Generalized Gradient Approximation.  These methods are very widely
applied in periodic boundary conditions calculations and are
computationally inexpensive, while retaining a realistic description
of the system.  Within an independent particle theory, such as
Kohn--Sham DFT, calculations for periodic systems are normally
performed by solving within the primitive unit cell.~\cite{note1} To
obtain the correct result for the infinite system it is necessary to
integrate over k--space, and the integral is normally approximated by
a sum over a finite set of k--points. A determinant formed from the
occupied orbitals at a single k--point in the first Brillouin zone
(BZ) of the primitive unit cell is a many--body wave function for a
simulation cell of the size of the primitive unit cell.  Adding a
second k--point doubles the size of the determinant and is equivalent
to doubling the size of the simulation cell, etc.

An important subtlety arises when finite size corrections derived from
an independent particle theory such as LDA--DFT are used to correct
the results of a true many--body method such as QMC.  Suppose the
many--body calculation is performed using the Ewald interaction, so
that the Coulomb energy is that of an infinite periodic array of
copies of the simulation cell.  Given that the solid we are trying to
model is crystalline, and hence the charge density is truly periodic,
the Ewald interaction gives a good description of the classical
Coulomb or Hartree energy.  However, because the electronic positions
are mirrored exactly in every copy of the simulation cell, the
electronic correlations are also forced to be periodic, and the
exchange--correlation (XC) energy corresponds to a system with a
periodic XC hole.  This unphysical approximation is particularly
inaccurate when the simulation cell is small.  In Kohn--Sham DFT
calculations the XC energy is evaluated using a standard functional
such as the LDA of Perdew and Zunger~\cite{PZ}, which was obtained by
fitting to the results of DMC calculations for jellium.~\cite{CA}
There is an important difference, however, in that the DMC results
were extrapolated to the infinite--system--size limit before the fit
was made.

The consequences of this difference are nicely illustrated by
considering many--body and LDA calculations for jellium, in which the
charge density is uniform.  The LDA gives the XC energy for the
infinite system irrespective of the size of the simulation cell, but
the XC energy obtained from a many--body simulation using the
Hamiltonian of Eq.~\ref{hamiltonian} with the Ewald interaction gives
an XC energy which depends on the size and shape of the simulation
cell.  Consequently, evaluating the correction term in
Eq.~\ref{fs_corr-extrap} within the LDA does not give a good
approximation to the finite size correction in the many--body
simulation and a significant extrapolation term remains.

The issue of finite size corrections to both the kinetic and
interaction energies has been addressed by Ceperley and
coworkers.~\cite{cep78,dmc87,tancep,kwon} Their approach is to add
separate extrapolation terms for the kinetic and interaction energies.
In their work on hydrogen solids, Ceperley and Alder~\cite{dmc87}
performed DMC calculations for a number of different system sizes and
fitted to the formula

\begin{equation}\label{cep_corr}
E^{\rm DMC}_{\infty} \simeq E^{\rm DMC}_N + a(T_{\infty} - T_{N}) +
\frac{b}{N}\;\;,
\end{equation}

\noindent where $a$ and $b$ are parameters, and $T$ is the kinetic
energy of the non--interacting electron gas.  The $b/N$ term accounts
for the finite size effects arising from the interaction energy and
the difference of the parameter $a$ from unity accounts for the
difference between the kinetic energies of the interacting and
non--interacting systems.  Normally $a>1$ because the interacting
kinetic energy is larger than the non--interacting kinetic energy.
Engel {\it et al.}~\cite{engel} used the following formula for
inhomogeneous systems:

\begin{equation}
E^{\rm QMC}_{\infty} \simeq E^{\rm QMC}_N + a(E^{\rm LDA}_{\infty} -
E^{\rm LDA}_{N}) + \frac{b}{N}\;\;,
\end{equation}

\noindent which reduces to Eq.~\ref{cep_corr} for a homogeneous
system.

As mentioned above, a large part of the finite size error in the
interaction energy arises from the use of the Ewald
interaction.~\cite{lmf96,ajw97} Our approach to the problem of the
finite size errors differs from that of Ceperley {\it et al.} in an
important respect.  We try to reduce the finite size effects within
the {\it many--body} calculation by modifying the interaction terms in
the many--body Hamiltonian.  The Ewald interaction is a periodic
function which differs from $1/r$ in such a way that the sum of
interactions between pairs of particles \emph{within one cell} gives
the exact Coulomb energy per cell of a periodic lattice of identical
cells.  This ensures that the Ewald Hamiltonian gives the correct
Hartree energy, but the deviations from $1/r$ give rise to a spurious
contribution to the XC energy, corresponding to the periodic
repetition of the XC hole discussed earlier.~\cite{lmf96,ajw97} Our
solution is to change the many--body Hamiltonian so that the
interaction with the XC hole is exactly $1/r$ (see
section~\ref{sec:cfse}) for any size and shape of simulation cell,
without altering the form of the Hartree energy.  This source of
finite size error is therefore eliminated and Eq.~\ref{fscorrection},
with the correction term calculated within the LDA, gives a much
better description of the remaining finite size errors.  Greater
accuracy can be obtained by adding an extrapolation term, but this
term is much smaller than if the Ewald interaction is used.

The idea of changing the Hamiltonian to reduce the finite size errors
may seem strange at first, so it is worth discussing a simple example.
Imagine that a periodic boundary condition QMC technique is being used
to study an isolated molecule.  If the molecule is placed at the
center of the periodically repeated simulation cell, the calculated
energy is that of an array of identical molecules, including unwanted
inter--molecular interactions.  The results improve as the simulation
cell is made larger, but the convergence is slow, especially for
molecules with permanent dipole moments.  A better solution is to cut
off all Coulomb interactions between charges on different molecules,
i.e., to replace the Ewald interactions by truncated Coulomb $1/r$
interactions acting only within the simulation cell.  As long as the
molecular wave function has decayed to zero before the simulation cell
boundary is reached, this procedure should give essentially exact
results.  The changes to the interaction to be discussed in this paper
are a generalization of this approach to make it useful in simulating
genuinely periodic systems such as crystals.  In this case the very
``strict'' notion of periodicity built into the Ewald interaction
produces an artificial periodic replication of the XC hole as well as
the required periodic replication of the charge density.  Our modified
Hamiltonian removes the effects of the unwanted extra periodicity just
as the simple truncation removed the unwanted periodicity in the
molecular example.

An alternative procedure is to evaluate the correction term in
Eq.~\ref{fscorrection} using HF data.  HF theory is an approximate
method for solving the many--body Hamiltonian, and if we use the Ewald
formula for the electron--electron interaction terms in both the
many--body and HF theories, the finite size error in the HF exchange
energy will tend to cancel the finite size error in the many--body
interaction energy.  At first sight this appears to be an excellent
solution to the finite size problem.  However, this procedure gives
too large a correction, presumably because the HF exchange hole is
significantly different from the screened XC hole of the many--body
system.

\section{Independent Particle Finite Size effects}
\label{sec:ipsereduction}

We call the correction term $(E_{\infty}^{\rm LDA} - E_{N}^{\rm LDA})$
the ``independent particle finite size effect'' (IPFSE). The finite
size effect arising from the particular model interaction used is
called the ``Coulomb finite size effect'' (CFSE).  In this section we
discuss the IPFSE in more detail.  The ``finite size error'' in an LDA
calculation for a perfect crystal arises from errors in the BZ
integration.  In an LDA calculation the computational cost is
proportional to the number of k--points used.  However, in a QMC
calculation the volume of the simulation cell is proportional to the
number of k--points in the primitive BZ and the computational cost
increases approximately as the cube of the volume of the simulation
cell. This means that it is even more important to choose the
k--points carefully in a QMC calculation so as to make the IPFSE as
small as possible.

Additional errors arise when using the supercell approximation for
systems which break translational symmetry.  Consider a crystal
containing a point defect.  In the supercell approach a finite
simulation cell containing a single defect is repeated throughout
space to form an infinite three--dimensional array.  The supercell
must be sufficiently large that the interaction between defects in
different cells is negligible.  This type of finite size effect is
distinct from the BZ integration error because it persists even when
the BZ integration is performed exactly.  However, the correction to
the defect formation energy from this type of finite size error should
be described reasonably well by a $(E_{\infty}^{\rm LDA} - E_{N}^{\rm
LDA})$ correction term, where $E_{\infty}^{\rm LDA}$ and $E_{N}^{\rm
LDA}$ are, respectively, the defect formation energies for a large
simulation cell and a smaller one, each containing a single defect.

A determinantal Bloch wave function suitable for use in a QMC
calculation may be formed from a set of single particle orbitals at a
single k--point, ${\bf k}_{\rm s}$, in the Brillouin zone of the
simulation cell reciprocal lattice.  The IPFSE can be greatly reduced
for insulating systems by a careful choice of ${\bf k}_{\rm s}$ using
the method of ``special k--points'' borrowed from band structure
theory.~\cite{baldereschi,MP} We have described the theory of the
``special k--points'' for many--body calculations in our earlier
work~\cite{gr94,gr95} and here we concentrate on the practicalities of
choosing ${\bf k}_{\rm s}$.  Baldereschi~\cite{baldereschi} defined
the ``mean--value point'', which is a k--point at which smooth
periodic functions of wave vector accurately approximate their
averages over the BZ, and clearly the Baldereschi mean--value point is
a strong candidate for ${\bf k}_{\rm s}$.  However, we find it
convenient to choose ${\bf k}_{\rm s}$ to be equal to half a
translation vector of the simulation cell reciprocal lattice (${\bf
k}_{\rm s}={\bf G}/2$), which allows the construction of real
single--particle orbitals and hence is computationally more efficient.
Some freedom is still left in the selection of ${\bf k}_{\rm s}$, and
we choose from the ${\bf G}/2$ according to the symmetrized plane wave
test of BZ integration quality introduced by
Baldereschi.~\cite{baldereschi} For example, for an fcc simulation
cell the half--reciprocal--lattice vectors correspond to the $\Gamma$,
$X$ and $L$ points of the BZ of the simulation cell reciprocal
lattice.  For a crystal with the full cubic symmetry ${\bf k}_{\rm s}
\equiv L$ gives the best BZ integration and ${\bf k}_{\rm s} \equiv
\Gamma$ the worst.

It is illuminating to relate these k--point schemes to the
multi--point schemes used in LDA and HF calculations.  In LDA and HF
calculations one normally samples the BZ of the primitive unit cell,
whereas in a QMC calculation one samples the BZ of the simulation
cell.  Suppose that the simulation cell has translation vectors $N_1
{\bf a}_1$, $N_2 {\bf a}_2$, and $N_3 {\bf a}_3$, where the $N_i$ are
integers and the ${\bf a}_i$ are vectors defining the primitive unit
cell.  When unfolded into the BZ of the primitive unit cell the
k--point ${\bf k}_{\rm s}$ maps onto the regular mesh

\begin{equation}
\label{kmesh}
{\bf k}_{lmn} = \frac{l}{N_1}{\bf b}_1 + \frac{m}{N_2}{\bf b}_2 +
\frac{n}{N_3}{\bf b}_3 + {\bf k}_{\rm s}\;\;\; ,
\end{equation}

\noindent where $l=0,\dots N_1-1, m=0,\dots N_2-1, n=0,\dots N_3-1$,
and the ${\bf b}_i$ are reciprocal to the ${\bf a}_i$.  This mesh is
of the same type as defined in the widely--used Monkhorst--Pack
scheme~\cite{MP} for BZ integration, differing only by an offset from
the origin.  The multi--point generalization of the Baldereschi scheme
which can be used in LDA and HF calculations is now obvious: one
chooses the offset ${\bf k}_{\rm s}$ to be equal to the mean--value
point~\cite{baldereschi} of the lattice defined by the translation
vectors $N_i {\bf a}_i$.  As mentioned above, for our purposes we
prefer to choose ${\bf k}_{\rm s}$ to be equal to half a translation
vector of the simulation cell reciprocal lattice.  This choice gives a
smooth and rapid convergence of the BZ integration with increasing
values of the $N_i$~\cite{gr95}.  For example, as mentioned above, for
fcc crystals we choose ${\bf k}_{\rm s}$ to be an $L$--point of the BZ
of the lattice defined by the $N_i {\bf a}_i$.  In this case the
Monkhorst--Pack~\cite{MP} definition of the offset corresponds to
taking an $L$--point for $N_i$ even, but the $\Gamma$--point for $N_i$
odd.  The latter choice gives poor results and this problem has
prompted some researchers to avoid Monkhorst--Pack meshes with odd
values of the $N_i$, but in fact odd values can be very efficient if
the mesh is offset according to our prescription.

\section{Coulomb Finite Size Effects}
\label{sec:cfse}

In this section we summarize the theory of the MPC interaction.  More
details and background are given in Refs.~\onlinecite{lmf96}
and~\onlinecite{ajw97}.  In section~\ref{sec:fscorr} we described how
the CFSE arises from the XC energy and the dependence of the Ewald
interaction, $v_{\rm E}$, on the size and shape of the simulation
cell.  Expanding $v_{\rm E}$ around zero separation
gives~\cite{lmf96,ajw97}

\begin{eqnarray}
\label{v_E}
v_{\rm E}({\bf r}) & = & {\rm constant} + \frac{1}{r} + \frac{2
\pi}{3\Omega} \; {\bf r}^{T}\cdot{\bf D}\cdot{\bf r} + {\cal
O}\left(\frac{r^4}{\Omega^{5/3}}\right) \;\;,
\end{eqnarray}

\noindent where $\Omega$ is the volume of the simulation cell, and the
tensor {\bf D} depends on the shape of the simulation cell. (For a
cubic cell {\bf D}={\bf I}.)  The constant term arises from the
condition that the average of $v_{\rm E}$ over the simulation cell is
zero. The term of order $r^2$ and the higher order deviations from
$1/r$ make the Ewald interaction periodic and ensure that the sum of
interactions between the particles in the simulation cell gives the
potential energy per cell of an infinite periodic lattice.  These
terms are responsible for the spurious contribution to the XC energy,
which is the source of the large finite size effect in many--body
calculations using the Ewald interaction.~\cite{lmf96,ajw97} For cubic
cells the interaction at short distances is larger than $1/r$ and
therefore the XC energy is more negative than it should be, and
because the leading order correction is proportional to the inverse of
the simulation cell volume the error per electron is inversely
proportional to the number of electrons in the cell.

Clearly it is desirable to remove this spurious contribution to the XC
energy, but we must remember that the Hartree energy is correctly
evaluated using the Ewald interaction.  The key requirements for a
model Coulomb interaction giving small CFSEs in simulations with
periodic boundary conditions are therefore: (i) it should give the
Ewald interaction for the Hartree terms, and (ii) it should be exactly
$1/r$ for the interaction with the XC hole.  Unfortunately, the only
periodic solution of Poisson's equation for a periodic array of
charges is the Ewald interaction, which obeys criterion (ii) only in
the limit of an infinitely large simulation cell.  We therefore
abandon the use of Poisson's equation and the Ewald interaction, and
instead use a ``Model Periodic Coulomb'' (MPC) interaction which
satifies both criteria.  This may seem a drastic step, but we point
out that we are trying to model an infinite system by a finite one and
that the inter--particle interaction used in the finite simulation
cell must model the effects of all the charges in the infinite system.

Our MPC interaction~\cite{ajw97} can be written as

\begin{eqnarray}
\lefteqn{ \hat{H}_{\rm e-e} = \sum_{i>j}f({\bf r}_i-{\bf r}_j)}
\nonumber \\ & + & \sum_{i}\int_{\rm WS} \left[ v_{{\rm E}}({\bf
r}_i-{\bf r})- f({\bf r}_i-{\bf r})\right] {\rho({\bf r})} \, {\rm
d}{\bf r} \;\;,
\label{eqcutoffhamil}
\end{eqnarray}
\noindent where $\rho$ is the electronic charge density and 
\begin{equation}
f({\bf r}) = \frac{1}{r_{\rm m}} \;\;.
\label{eqfofr}
\end{equation}

\noindent The definition of the cutoff Coulomb function, $f$, involves
a minimum image convention whereby the inter--electron distance, ${\bf
r}$, is reduced into the Wigner--Seitz (WS) cell of the simulation
cell lattice by removal of simulation cell lattice translation
vectors, leaving a vector ${\bf r}_{\rm m}$.  This ensures that
$\hat{H}_{\rm e-e}$ has the correct translational and point group
symmetry.  The first term in Eq.~\ref{eqcutoffhamil} is a direct
Coulomb interaction between electrons within the simulation cell and
the second term is a sum of potentials due to electrons ``outside the
simulation cell''.  Note that the second term is a one--body potential
similar to the Hartree potential.  It depends on the electronic charge
density, $\rho$, but is not a function of the separation of the
electrons.

The electron--electron contribution to the total energy is evaluated
as the expectation value of $\hat{H}_{\rm e-e}$ with the many-electron
wave function, $\phi$, minus a double counting term:

\begin{eqnarray}
E_{\rm e-e}& = &\langle \phi | \hat{H}_{\rm e-e} | \phi \rangle
\nonumber \\ & - & \frac{1}{2}\int_{\rm WS} \rho({\bf r})\rho({\bf r}') \left[ v_{{\rm
E}}({\bf r}-{\bf r}')- f({\bf r}-{\bf r}')\right] \, {\rm d}{\bf r} \,
{\rm d}{\bf r}' \;\;.
\label{eqcutoffE1}
\end{eqnarray}

\noindent Evaluating the expectation value gives

\begin{eqnarray}
E_{\rm e-e} & = & \frac{1}{2}\int_{\rm WS} \rho({\bf r})\rho({\bf r}')
v_{\rm E}({\bf r}-{\bf r}') \, {\rm d}{\bf r} \, {\rm d}{\bf r}'
\nonumber \\ & + & \left( \int_{\rm WS} |\phi|^2\sum_{i>j}f({\bf
r}_i-{\bf r}_j) \, \Pi_k \, {\rm d}{\bf r}_k  \right. \nonumber \\ &-&
\left. \frac{1}{2}\int_{\rm WS} \rho({\bf r})\rho({\bf r}') {f}({\bf r}-{\bf r}') \, {\rm d}{\bf
r} \, {\rm d}{\bf r}' \right) \;\; ,
\label{eqcutoffE2}
\end{eqnarray}

\noindent where the first term on the right hand side is the Hartree
energy and the term in brackets is the XC energy.  We can see
immediately that the Hartree energy is calculated with the Ewald
interaction while the XC energy (expressed as the difference between a
full Coulomb term and a Hartree term) is calculated with the cutoff
interaction $f$.

The charge density $\rho$ appears in Eqs.~\ref{eqcutoffhamil},
\ref{eqcutoffE1}, and \ref{eqcutoffE2}.  In QMC methods, the charge
density is known with greatest statistical accuracy at the end of the
calculation.  This is not a serious complication for VMC simulations
as the interaction energy may be evaluated at the end of the
simulation using the accumulated charge density.  In DMC this is not
possible because the local energy~\cite{dmc,hammond} is required at
every step.  We investigate this point further in
section~\ref{sec:dmcapp}.

The CFSE may be viewed in another way~\cite{lmf96}.  Consider a large
cluster of identical simulation cells.  Almost all possible
configurations of the electrons within the simulation cell have a net
dipole moment.  The dipole moments in the periodic replicas of the
simulation cell are aligned, resulting in a depolarization field
across the sample.  (In real non-ferroelectric solids the dipoles from
different regions are not aligned and the depolarization field is
greatly reduced.)  To make the electrostatic potential periodic deep
within the cluster we need to apply a cancelling external electric
field.  In the limit of a very large cluster this is equivalent to
using the Ewald interaction.  In a cubic system the total energy of
interaction between the dipoles, which includes the interaction with
the depolarization field, does not contribute to the ${\cal O}(r^2)$
term in Eq.~\ref{v_E}~\cite{kittel}.  For a cubic system the ${\cal
O}(r^2)$ term in Eq.~\ref{v_E} therefore arises solely from the
interaction between the dipoles and the external field.  In a
non--cubic system the ${\cal O}(r^2)$ term contains contributions from
both the dipole--dipole interactions and the interactions of the
dipoles with the external field.  The advantage of our MPC interaction
is that it prevents the ${\cal O}(r^2)$ terms contributing to the XC
energy.

If the simulation cell is not big enough to accommodate the true XC
hole then the XC hole is squeezed into the cell and a finite size
error ensues.  This appears to be the largest finite size error in the
interaction energy after the CFSE has been removed.  Extrapolation
currently appears to be the best method of correcting this error.
When the XC hole is squeezed by the finite simulation cell the XC
energy calculated using the MPC interaction is too negative.  The
effect of the ${\cal O}(r^2)$ term in the Ewald interaction is to make
it even more negative, however, and so the MPC interaction is still
better than the Ewald interaction.

\subsection{Systems of electrons and nuclei}
\label{sub:sysen}
In the previous sections we considered the electron--electron
interaction only, and did not discuss the electron--nucleus and
nucleus--nucleus interactions.  If the MPC interaction is physically
reasonable we should, however, be able to apply it to all the
interactions in the problem, not just the electron--electron
terms. Here we apply the MPC interaction to a system of electrons and
nuclei, showing that under certain common conditions the expressions
simplify so that the electron--nucleus and nucleus--nucleus terms
reduce to the Ewald form.

Consider a simulation cell with periodic boundary conditions
containing $N$ electrons at positions ${\bf r}_i$ and $M$ nuclei of
charge $Z_{\alpha}$ at positions ${\bf x}_{\alpha}$.  The wave
function of the electrons and nuclei is $\Psi(\{{\bf r}_i\},\{{\bf
x}_{\alpha}\})$, and the total charge density (electrons and nuclei)
is $\rho_{\rm T}({\bf r})$.  The interaction energy calculated with
the MPC interaction is

\begin{eqnarray}
\label{E_int}
E_{\rm int} & = & \frac{1}{2}\int_{\rm WS} \rho_{\rm T}({\bf
r})\rho_{\rm T}({\bf r}') [v_{{\rm E}}({\bf r}-{\bf r}') - f({\bf
r}-{\bf r}')] \, {\rm d}{\bf r} \, {\rm d}{\bf r}' \nonumber \\ & + &
\int_{\rm WS} |\Psi|^2 \left[ \sum_{i>j}f({\bf r}_i-{\bf r}_j)
\right. - \sum_{i} \sum_{\alpha} Z_{\alpha}f({\bf r}_i-{\bf
x}_{\alpha}) \nonumber \\ & + &\left.
\sum_{\alpha>\beta}Z_{\alpha}Z_{\beta}f({\bf x}_{\alpha}-{\bf
x}_{\beta}) \right] \, \Pi_k \, {\rm d}{\bf r}_k \, \Pi_{\gamma} \,
{\rm d}{\bf x}_{\gamma} \;\;\; .
\label{eq11}
\end{eqnarray}

We now employ the adiabatic approximation to separate the electronic
and nuclear dynamical variables:

\begin{equation}
\Psi (\{{\bf r}_i\},\{{\bf x}_{\alpha}\}) = \phi (\{{\bf r}_i\};\{{\bf
x}_{\alpha}\}) \Phi (\{{\bf x}_{\alpha}\}) \;\;\; ,
\label{eqsepelecion}
\end{equation}

\noindent where the $\{{\bf x}_{\alpha}\}$ appear only as parameters
in $\phi$.  To make further progress we must assume a form for the
nuclear part of the wave function, $\Phi$.  The simplest assumption is
that $\Phi$ can be written as an appropriately symmetrized product of
very strongly localized non--overlapping single--nucleus functions.
Regardless of whether the product is antisymmetrized, symmetrized, or
not symmetrized, Eq.~\ref{E_int} reduces to

\begin{eqnarray}
E_{\rm int} & = & \frac{1}{2}\int_{\rm WS} {\rho} ({\bf r}){\rho}
({\bf r}') [v_{{\rm E}}({\bf r}-{\bf r}') - f({\bf r}-{\bf r}')] \,
{\rm d}{\bf r} \, {\rm d}{\bf r}' \nonumber \\ & + & \int_{\rm WS}
|\phi|^2 \sum_{i>j}f({\bf r}_i-{\bf r}_j) \, \Pi_k \, {\rm d}{\bf r}_k
\nonumber \\ & - & \int_{\rm WS} |\phi|^2 \sum_{i}\sum_{\alpha}
Z_{\alpha} \, v_{{\rm E}}({\bf r}_i-\overline{{\bf x}}_{\alpha}) \,
\Pi_k \, {\rm d}{\bf r}_k \nonumber \\ & + &
\sum_{\alpha>\beta}Z_{\alpha}Z_{\beta} \, v_{{\rm E}}(\overline{{\bf
x}}_{\alpha}-\overline{{\bf x}}_{\beta}) \;\;\;,
\label{eqesystemtotal}
\end{eqnarray}

\noindent where the $\overline{{\bf x}}_{\alpha}$ denote the centers
of the single--nucleus functions and ${\rho}$ is the electron density.
Note that the electron--nucleus and nucleus--nucleus terms involve
only the Ewald interaction and that the first two terms of
Eq.~\ref{eqesystemtotal} correspond precisely to the
electron--electron interaction of Eq.~\ref{eqcutoffE2}.  This result
justifies the use of Eq.~\ref{eqcutoffhamil} for the
electron--electron interactions while retaining the Ewald interaction
for the electron--nucleus and nucleus--nucleus terms.

\section{Tests of the MPC interaction}

\subsection{Application to HF calculations}
\label{sec:hfapp}

We have tested the MPC interaction by performing a series of
calculations on diamond structure silicon using fcc simulation cells
whose translation vectors are $n$ times those of the primitive unit
cell.  In a previous publication we gave a few results from such
tests~\cite{ajw97}, but here we present new tests and subject them to
a more detailed analysis.

In our first set of tests we compare LDA and HF results.  These tests
are inexpensive, which allows us to study very large systems, and they
are not subject to statistical errors because Monte Carlo techniques
are not involved.  We consider simulation cells with $n$ = 1, 2, 3, 4,
and 5, which contain 2, 16, 54, 128, and 250 ions, respectively. The
Si$^{4+}$ ions were represented by norm--conserving non--local LDA
pseudopotentials and the calculations were performed using a
plane--wave basis set with a cutoff energy of 15 Ry.  To facilitate
comparison we evaluate the HF energy with the LDA orbitals, so that
the energy differences arise solely from the difference between the
LDA XC energy and the HF exchange energy. We performed calculations
using ${\bf k}_s \equiv L$ and ${\bf k}_s \equiv \Gamma$.  The HF
energy evaluated with the MPC interaction is

\begin{eqnarray}
E^{\rm HF} & = & \sum_i -\frac{1}{2}\int \phi_i^*({\bf r}) \nabla^2
\phi_i({\bf r}) \, {\rm d}{\bf r} \nonumber \\ & + & \frac{1}{2} \int
\rho({\bf r}) \rho({\bf r}') v_{\rm E}({\bf r}-{\bf r}') \, {\rm
d}{\bf r} \, {\rm d}{\bf r}' \nonumber \\ & - & \frac{1}{2} \sum_{i,
j}^N \delta_{s_is_j} \!\! \int \phi_i^*({\bf r}) \phi_i({\bf r}')
f({\bf r}-{\bf r}') \phi_j^*({\bf r}') \phi_j({\bf r}) \, {\rm d}{\bf
r} \, {\rm d}{\bf r}' \nonumber \\ & + & \int V_{\rm ext}({\bf r})
\rho({\bf r}) \, {\rm d}{\bf r} + E_{\rm I-I} \;\;\; ,
\label{HFenergy}
\end{eqnarray}

\noindent where $E_{\rm I-I}$ is the ion--ion energy calculated with
the Ewald interaction.  Note that the Hartree energy is evaluated with
the Ewald interaction while the exchange energy is evaluated with the
cutoff Coulomb interaction.

In Fig.~\ref{hfldadataG} we show the deviations of the LDA and HF
energies from the fully converged values as a function of system size
for ${\bf k}_s \equiv \Gamma$ wave functions. The LDA energy converges
smoothly with system size but for small system sizes the IPFSE error
is large because of the $\Gamma$--point sampling.  We show HF energies
calculated with the Ewald and MPC interactions, with and without
incorporating the IPFSE corrections obtained from the LDA data.  The
data incorporating the IPFSE (filled symbols) show the residual CFSE
errors. The IPFSE is positive while the CFSE is negative, in accord
with the analysis of the CFSE given in section~\ref{sec:cfse}. The
IPFSE corrected data show that the CFSE for the MPC interaction is
roughly half that for the Ewald interaction.
 
In Fig.~\ref{hfldadataL} we show similar data for ${\bf k}_s \equiv
L$.  The LDA energy converges rapidly and smoothly with system size,
and therefore the IPFSE is small, which demonstrates the efficacy of
the ``special k--points'' method.  The IPFSE and CFSE errors tend to
cancel and for ${\bf k}_s \equiv L$ sampling the HF data converge more
rapidly without the IPFSE corrections. For $n$ = 3, corresponding to a
54 atom simulation cell, the LDA finite size error (IPFSE only) is
0.011 eV per atom, which is very much smaller than the HF (Ewald)
finite size error of -0.211 eV per atom or the equivalent HF (MPC)
error of -0.071 eV per atom.  As in the case of $\Gamma$--point
sampling, the CFSE for the MPC interaction is roughly half that for
the Ewald interaction.

After applying the IPFSE correction obtained from the LDA data the HF
results for ${\bf k}_s \equiv \Gamma$ and ${\bf k}_s \equiv L$
sampling are very similar.  The correspondence of the IPFSE corrected
data for the $L$-- and $\Gamma$--points demonstrates that to a very
good approximation the IPFSE and CFSE are independent.  Estimating the
energy of the infinite system by averaging the energy over a set of
${\bf k}_s$ vectors removes the IPFSE but does not remove the CFSE.

We have fitted the CFSE errors from the filled data points in
Figs.~\ref{hfldadataG} and~\ref{hfldadataL} to the form $b/N^x$, where
$b$ and $x$ are parameters and $N = 8n^3$ is the number of electrons
in the simulation cell.  The fits give values of $x$ in the region of
unity for both the Ewald and MPC interactions.  This extrapolation
function is therefore suitable for both interactions, although the
size of the extrapolation term is smaller for the MPC interaction.

The Ewald and MPC interactions differ by an amount inversely
proportional to the volume of the simulation cell.  This means that
although the energies per particle converge to the same value as the
volume of the simulation cell increases, the difference between the
Ewald and MPC energies of the whole simulation cell tends to a finite
constant as the simulation cell volume goes to infinity.  We have
evaluated this constant value for the fixed--LDA--orbital HF
calculations described in this section by extrapolating the energy
difference between the Ewald and MPC energies.  This gives a value of
approximately 14 eV.

\subsection{Application to VMC}
\label{sec:vmcapp}

In the VMC method~\cite{vmc,hammond} the energy is calculated as the
expectation value of the Hamiltonian, $\hat{H}$, with a trial wave
function, $\phi_{\rm T}$, yielding a rigorous upper bound to the exact
ground state energy.  The Metropolis algorithm is used to generate
electron configurations distributed according to $\phi_{\rm T}^2$, and
the energy calculation is performed by averaging the local energy,
$\phi_{\rm T}^{-1} \hat{H} \phi_{\rm T}$, over this distribution.

Our trial wave functions are of the standard Slater--Jastrow type:

\begin{equation}\label{hfjchi_trial}
\phi_{\rm T} = D^{\uparrow} D^{\downarrow} \exp \left[ \sum_{i=1}^{N}
\chi({\bf r}_i) - \sum_{i<j}^{N} u(r_{ij}) \right] \;\;\; ,
\end{equation}

\noindent where there are $N$ electrons in the simulation cell, $\chi$
is a one--body function, $u$ is a two--body correlation factor which
depends on the relative spins of the two electrons, and $D^{\uparrow}$
and $D^{\downarrow}$ are Slater determinants of up-- and down--spin
single--particle orbitals.  The $u$ functions were of the type
described in~\onlinecite{opt_prb}, while for the $\chi$ functions we
used spherically symmetric functions centered on each atom.  These
$\chi$ functions give significantly better results than the truncated
Fourier series representation used in our earlier
work.~\cite{opt_prb,ajw97} The trial wave functions contained 32
variable parameters, whose optimal values were obtained by minimizing
the variance of the energy using 10,000--20,000 statistically
independent electron configurations, which were regenerated several
times during the minimization procedure.~\cite{opt_prb,umrigar} We
used the same pseudopotential as in our LDA and HF calculations,
sampling the non--local potential using the techniques of Fahy {\it et
al.}~\cite{fahy_prb}.

We have optimized wave functions using both the Ewald and MPC
interactions.  The wave functions generated using the different
interactions are almost identical.  Properties other than the energy,
such as pair correlation functions, are therefore hardly affected by
the choice of interaction.  As the MPC interaction gives the correct
interaction between the electrons at short distances it may give a
better account of, for example, the short distance behaviour of the
pair correlation function, but more numerical work is required to
investigate this point.

In Fig.~\ref{vmcdata} we show results for VMC calculations of the
energies of the same systems as in the previous section.  These
results are similar to those given in Ref.~\cite{ajw97}, but they have
been recalculated for this work with more accurate wave functions and
better sampling, and the data have been corrected for the IPFSE.  The
total energies were calculated to a statistical accuracy of $\pm 0.01$
eV per atom.  The VMC data display a smaller CFSE than the HF data,
probably because the HF exchange hole is longer ranged than the
screened hole obtained in the correlated calculations.  For $n$ = 2,
the MPC interaction reduces the VMC finite size error by more than
50\%, from $-0.403$ to $-0.187$ eV per atom.

\subsection{Application to DMC}
\label{sec:dmcapp}

In the DMC method \cite{dmc,hammond}, imaginary time evolution of the
Schr\"{o}dinger equation is used to evolve an ensemble of
3$N$--dimensional electronic configurations towards the ground state.
The calculations are made tractable by using the fixed node
approximation and by incorporating importance sampling.  The method
generates the distribution $\phi_{\rm T}\psi$, where $\psi$ is the
best (lowest energy) wave function with the same nodes as the guiding
wave function, $\phi_{\rm T}$.  The accuracy of the fixed node
approximation can be tested on small systems and the results are
normally very satisfactory.\cite{hammond}

We evaluated the non--local pseudopotential energy using the
``locality approximation''.~\cite{nl_dmc} The short--time
approximation for the Green's function was used with a time step of
0.01 a.u.  Li {\it et al.}~\cite{li} found that using a time step of
0.015 a.u. gave a time--step error of less than 0.03 eV per atom in
silicon, so our time step error should be even smaller. The total
energies were calculated to a statistical accuracy of $\pm 0.02$ eV
per atom.

Our MPC interaction is more complicated to apply in DMC than in VMC
because the evaluation of the importance sampled Green's function
requires the local energy.  The modified Hamiltonian, and hence the
local energy, depends on the charge density, and therefore we must
know the charge density before we can perform the DMC calculation.
Fortunately, however, the local energy is relatively insensitive to
the charge density used in the Hamiltonian (Eq. \ref{eqcutoffhamil})
because $v_{\rm E}({\bf r})-f({\bf r})$ is small when $|{\bf r}|$ is
small.

We have tested the sensitivity of the Green's function to the charge
density used in the Hamiltonian. Two candidate charge densities are
the charge density of the determinantal part of the QMC wave function
and the charge density of the VMC guiding wave function.  Even for a
small system ($n$ = 2) we find it sufficient to use the LDA charge
density during the calculation of the Green's function and to
re--evaluate the charge density dependent term in the interaction
energy using the DMC density obtained at the end of the calculation.
The sensitivity rapidly reduces with increasing system size, and this
procedure gives errors of less than 0.03 eV per atom for $n$ = 2, and
less than 0.01 eV per atom for larger system sizes.  Therefore, the
requirement of having a good approximation to the charge density in
advance of the DMC calculation does not pose a significant difficulty.
A successful DMC calculation requires a good quality VMC trial
function and its charge density can be obtained during the process of
wave function optimization.

In Fig.~\ref{dmcdata} we show results for DMC calculations on the same
systems as for our VMC study, the largest of which contains 1000
electrons.  The results include a correction for the IPFSE.  The
convergence behaviour is very similar to the VMC data.  The MPC
energies are always above those for the Ewald interaction and the MPC
interaction significantly reduces the CFSE.  These results demonstrate
that the finite size errors within VMC and DMC calculations are very
similar and that the MPC interaction is similarly effective in both
methods.

We have fitted the residual CFSE errors in both the VMC and DMC
calculations to the form $b/N^x$.  The fit is reasonable and gives a
value of $x$ close to unity for both the Ewald and MPC data.  For
these data it is possible to obtain more accurate approximations to
$E_{\infty}^{\rm QMC}$ by using such an extrapolation function,
although the calculations for the large system sizes are costly,
especially within DMC.  The extrapolated energies should be more
accurate for the MPC interaction because the extrapolation term is
significantly smaller.

Many interesting applications of VMC and DMC methods will be to
problems in which the quantity of physical interest is the difference
in energy between two large systems.  Examples of such problems are
calculations of excitation energies and defect energies in solids.  In
such cases the energy of interest is approximately independent of the
size of the simulation cell, so that for each simulation cell size it
is the energy of the {\it whole simulation cell} which must be
converged to the required tolerance, not the energy per atom as we
plotted in Figs.~\ref{vmcdata} and~\ref{dmcdata}.  In these cases
extrapolation will be so costly that it can hardly be contemplated.
In some cases the CFSE will largely cancel between the two systems, as
occurs in our excitation energy calculations (see next section).  This
cancellation cannot always be relied upon, however, especially when
the simulation cells contain different numbers of particles, and the
use of the MPC interaction should be particularly beneficial in such
cases.

\section{Excitation energies}
\label{sec:exen}

The quasiparticle excitation energies are the energies for adding an
electron to the system or subtracting one from it.  A quasiparticle
energy has both real and imaginary parts, the latter giving the
quasiparticle lifetime.  For the minimum energy electron and hole
quasiparticle excitations the imaginary parts of the quasiparticle
energies are zero and the quasiparticles have an infinite lifetime.
In this case the exact quasiparticle energy gap can be written as

\begin{equation}
E_{\rm g} = (E_{N+1} - E_{N}) + (E_{N-1} - E_{N})\;,
\end{equation}

\noindent where $E_{N+1}$, $E_{N-1}$, and $E_{N}$ are the ground state
total energies of the $N+1$, $N-1$ and $N$ electron systems.  A
general quasiparticle energy gap cannot be written in terms of
differences between energies of exact eigenstates of the system, but
such an approximation is often accurate for low energy gaps.  The
quasiparticle energies are measured in photoemission and inverse
photoemission experiments.  In an optical absorption experiment a
different process occurs in which an electron is excited from the
valence to the conduction band.  This introduces two quasiparticles
into the system, the electron and hole, which interact and can form an
exciton, in which case the lowest excitation energy is smaller than
$E_{\rm g}$ by the exciton binding energy, $E_{\rm b}$.

The HF method gives approximations to the energies of quasiparticles
and the interactions between them.  According to Koopmans' theorem, if
orbital relaxation is neglected, the quasiparticle energies are equal
to the HF eigenvalues.  Koopmans' theorem can be extended to include
correlation effects.~\cite{ekt1,ekt2} The extended Koopmans' theorem
has been used in conjunction with VMC methods to calculate
quasiparticle energies in silicon.~\cite{natural} In both of these
methods the ``quasiparticle energies'' are real as they are obtained
as approximations to the energy differences between exact eigenstates
of the system.

Recently there has been significant progress in applying QMC
techniques to calculate approximate excitation energies from
eigenstates using ``direct methods''.  In these approaches an
excitation energy is obtained by performing separate QMC calculations
for the ground and excited states.  A Slater--Jastrow wave function is
used for the ground state, and for the optical gap the excited state
is formed by replacing a valence band single particle orbital by a
conduction band one.  We call this a ``promotion'' calculation, and
such calculations have been reported for a nitrogen
solid~\cite{mitas1}, diamond~\cite{mitas2,mitas3}, and
silicon~\cite{bandgap}.  Photoemission/inverse photoemission gaps may
be obtained by using QMC to calculate the ground state energies of the
${N+1}$, ${N-1}$, and ${N}$ electron systems.  Wave functions for the
${N+1}$ and ${N-1}$ electron systems may be formed by adding or
subtracting an orbital from the up-- or down--spin determinants of the
$N$--electron wave function.  We call this an ``addition/subtraction''
calculation.  For calculations with periodic boundary conditions the
simulation cell is made charge neutral by adding a compensating
uniform background charge density.  Calculations of this type have
been reported for one--~\cite{knorr} and two--dimensional~\cite{engel}
model systems, while results for a three--dimensional system (silicon)
are reported in this paper.

QMC calculations of excitation energies in extended systems are
computationally very demanding because they are `$\frac{1}{N}$'
effects, i.e., the fractional change in energy is inversely
proportional to the number of electrons in the system.  This means
that high statistical accuracy is required to obtain good results.
The largest system for which excitation energies have been calculated
prior to this paper is 16 atoms (64 electrons).~\cite{bandgap} The
total finite size error in the ground state energy for that system was
estimated to be about 16 eV per simulation cell, while the energy
scale of interest for the excitations is of order 0.1 eV.  Like almost
all methods for calculating excitation energies, QMC calculations of
this type only work because of a strong cancellation of errors between
the ground and excited states.  It turns out that the finite size
errors tend to reduce the energy gap, while the errors in the trial
wave functions are usually larger for the excited states than for the
ground state and so increase the energy gap.  Although good agreement
with experimental excitation energies has been found using small
simulation cells,~\cite{mitas2,mitas3,bandgap} one is left with the
suspicion that if larger simulation cells were used the agreement with
experiment might be significantly worse because the finite size
effects would be smaller.  Before these QMC techniques can be relied
upon for calculating excitation energies it is necessary that the
issue of finite size effects be properly explored.  In the next
sections we address the following questions:

\begin{enumerate}
\item What are the sizes and origins of the finite size effects in
excitation energy calculations?
\item What are the differences in finite size effects between
promotion and addition/subtraction calculations?
\end{enumerate}

\subsection{HF theory of excitation energies}
\label{HFexcite}
First we consider excitation energies for solids within HF theory.
The HF equations with the MPC interaction are obtained by minimizing
the HF energy of Eq.~\ref{HFenergy} with respect to the
single--particle orbitals, giving

\begin{eqnarray}
& - &\frac{1}{2}\nabla^2 \phi_i + \int \rho({\bf r}') v_{\rm E}({\bf
r}-{\bf r}') \, {\rm d}{\bf r}' \, \phi_i \nonumber \\ & - &
\sum_{i,j}^N\delta_{s_is_j} \!\! \int \phi_j^*({\bf r}') \phi_j({\bf
r}) \phi_i({\bf r}') f({\bf r}-{\bf r}') \, {\rm d}{\bf r}' \nonumber \\& +& V_{\rm
ext} \phi_i = \epsilon_i \phi_i \;\;\; .
\label{HFequation}
\end{eqnarray}

\noindent If we neglect the relaxation of the orbitals, the energy
required to excite an electron from the $j^{th}$ (occupied) orbital
into the $i^{th}$ (unoccupied) orbital is

\begin{eqnarray}\label{HF_promote_gap}
\Delta E_{ij} & = & (\epsilon_{i} - \epsilon_{j}) - \int \rho_i({\bf
r}) \rho_j({\bf r}') v_{\rm E}({\bf r}-{\bf r}') \, {\rm d}{\bf r} \,
{\rm d}{\bf r}' \nonumber \\
& +& \delta_{s_is_j} \int \phi_i^*({\bf r}) \phi_j^*({\bf
r}') \phi_j({\bf r}) \phi_i({\bf r}') f({\bf r}-{\bf r}') \, {\rm
d}{\bf r} \, {\rm d}{\bf r}' \nonumber \\
& + & \frac{1}{2} \int \rho_i({\bf r}) \rho_i({\bf r}') \left[ v_{\rm E}({\bf r}-{\bf
r}')-f({\bf r}-{\bf r}')\right] \, {\rm d}{\bf r} \, {\rm d}{\bf r}'
\nonumber \\
& + & \frac{1}{2} \int \rho_j({\bf r}) \rho_j({\bf r}')
\left[ v_{\rm E}({\bf r}-{\bf r}')-f({\bf r}-{\bf r}')\right] \, {\rm
d}{\bf r} \, {\rm d}{\bf r}'\;,
\end{eqnarray}

\noindent where $\rho_k = |\phi_k|^2$ is the charge density from the
$k^{th}$ orbital.  The first term is the eigenvalue difference for the
excitation while the second and third terms are the Hartree and
exchange interactions between the electron and hole.  Within this
approximation the electron--hole terms go to zero in the limit of an
infinitely large simulation cell.  The fourth and fifth terms on the
right hand side are absent if one uses the Ewald interaction instead
of the MPC interaction, i.e., we replace $f$ by $v_{\rm E}$.  When the
relaxation of the orbitals is neglected these terms also go to zero
when the size of the simulation cell goes to infinity because $v_{\rm
E}$ tends to $1/r$ over most of the simulation cell.

The addition/subtraction gap is given by

\begin{eqnarray}
E_{\rm g}& =& \left(E_{+i}^{\rm HF}-E_0^{\rm HF}\right) - \left(E_0^{\rm
HF}-E_{-j}^{\rm HF}\right) \nonumber \\ & = & (\epsilon_i - \epsilon_j) \nonumber
\\ & + & \frac{1}{2} \int \rho_i({\bf r}) \rho_i({\bf r}') \left[
v_{\rm E}({\bf r}-{\bf r}')-f({\bf r}-{\bf r}')\right] \, {\rm d}{\bf
r} \, {\rm d}{\bf r}' \nonumber \\ & + & \frac{1}{2} \int \rho_j({\bf
r}) \rho_j({\bf r}') \left[ v_{\rm E}({\bf r}-{\bf r}')-f({\bf r}-{\bf
r}')\right] \, {\rm d}{\bf r} \, {\rm d}{\bf r}' \;\;,
\label{koopgap}
\end{eqnarray}

\noindent where $E_0^{\rm HF}$ is the HF ground state energy of the
$N$--electron system, $E_{+i}^{\rm HF}$ is the energy of the state
with an electron added to the $i^{th}$ (previously unoccupied)
orbital, along with the uniform background charge, and $E_{-j}^{\rm
HF}$ is the energy of the state where an electron is removed from the
$j^{th}$ orbital, along with the background charge.  The standard
Koopmans' theorem has been modified and contains two additional terms,
which also occur in the promotion energy, $\Delta E_{ij}$.  We have
evaluated these additional terms using LDA orbitals and have found
that even for a small simulation cell ($n$ = 2) they are very small,
being in the range $\pm$0.05 eV, and they decrease rapidly with system
size.  We do not expect that the use of exact HF orbitals or orbital
relaxation will greatly affect these results.

In Fig.~\ref{fig_koopgap} we show the addition/subtraction energies
calculated using Eq.~\ref{koopgap} for the $\Gamma_{25'} \!
\rightarrow \! \Gamma_{15}$ energy gap and the valence band width,
calculated with both the Ewald and MPC interactions, along with LDA
values.  (The results for other energies show similar behaviour.)  We
do not show the promotion energies in Fig.~\ref{fig_koopgap} because
they differ from the addition/subtraction energies only by the exciton
binding energy, which decreases with increasing system size quite
rapidly.  Fig.~\ref{fig_koopgap} shows that the HF results for the
Ewald and MPC interactions are very similar.  The band width converges
by about $n$ = 7, but the band gap is still slowly increasing at $n$ =
12, and the Ewald and MPC values are not yet equal, which they must be
at convergence.  For the largest system size studied ($n$ = 12) the
MPC gap and valence band width are 7.4 eV and 17.7 eV respectively,
which are a little smaller than the HF values of 8.0 eV and 18.9 eV
given in Ref.~\onlinecite{HF}.  Presumably the major reasons for these
differences are that we use LDA wave functions and LDA--derived
pseudopotentials, although as noted above there is clear evidence that
in our calculations the HF energy gap has still not fully converged at
$n=12$.  The LDA excitation energies converge very rapidly with system
size.  Note that this would not be true in either LDA or HF theory if
we studied isolated clusters of atoms.  In a recent study of silicon
clusters, \"{O}\v{g}\"{u}t {\it et al.}~\cite{ogut} found large
differences between the band gap in the LDA eigenvalues and the band
gap calculated by electron addition/subtraction.  As shown by
Franceschetti {\it et al.}~\cite{franceschetti}, these differences are
due to the charging of the cluster when an electron is added or
subtracted, which does not occur in our calculations because a uniform
background is added to preserve charge neutrality.  The slow
convergence of the HF excitation energies apparent in
Fig.~\ref{fig_koopgap} therefore arises from the exchange energy.
Moreover, because the results with the Ewald and MPC interactions are
almost the same, the source of the error is not the interaction with
the exchange hole, but the shape of the exchange hole.  This is
because the excitation energy depends on the change in the exchange
hole due to the excitation, which is not strongly localised.

In a very interesting set of calculations Engel {\it et
al.}~\cite{engel} studied excited states of a model two--dimensional
system using LDA, $GW$, VMC and DMC techniques.  They performed a
number of VMC calculations with increasing system size and found that
the addition/subtraction gap tended to increase with system size,
which is the same behaviour as we have found in our HF calculations.
Engel {\it et al.} went on to give an explanation of this effect.
Their explanation was that in the $N+1$ ($N-1$) electron systems there
is an additional electrostatic energy due to the interaction of the
extra electron (hole) with the additional electron (hole) in the other
simulation cells.  Taking into account the additional compensating
uniform background charge that was added to keep each cell neutral,
this additional energy is negative and therefore the energies
$E_{N+1}$ and $E_{N-1}$ are lower than they should be.  Engel {\it et
al.}  showed that the observed finite size effect is much smaller than
the Madelung energy for point charges, and to explain this they argued
that the effect would be screened by the response of the other
electrons.  This argument implies that the finite size effects in
addition/subtraction calculations are larger than those in promotion
calculations.

Our analysis of the situation is as follows.  For simplicity we
consider our HF calculations, where the interaction energy can be
divided into Hartree and exchange contributions.  The significant
underestimation of the HF band gaps of small systems is not due to the
Hartree terms, which by construction are the same as for our LDA
calculations and give very small finite size effects in the band
gaps. The finite size error in the HF gaps therefore arises from the
exchange energy.  By comparing band gaps calculated with the Ewald and
MPC interactions we can see whether the problem lies with the
interaction or with the shape of the exchange hole.  Because we find
that the band gaps calculated with the Ewald and MPC interactions are
very similar we conclude that the form of the interaction is not the
important consideration. Therefore the source of the problem must be
the finite size errors present in the shape of the exchange hole.
This argument implies that the finite size effects in
addition/subtraction calculations are similar to those in promotion
calculations.  Our viewpoint is supported by the HF results that have
been presented in this sub--section and also by the VMC results to be
discussed in the next sub--section.

In summary, the HF excitation energies calculated with the Ewald and
MPC interactions are very similar.  Within HF theory the largest
finite size error in excitation energies arises from the shape of the
exchange hole, which leads to slow convergence with system size. The
finite size errors in promotion and addition/subtraction HF
calculations are of similar size.

\subsection{QMC theory of excitation energies}
\label{QMCexcite}
We now apply the theory developed in the previous section to QMC
calculations of excitation energies.  Although we have just
demonstrated that the HF gaps converge rather slowly with system size,
we showed earlier that the finite size effects in VMC and DMC ground
state energies are smaller than in HF theory.  It is important to see
whether this also applies to excitation energies.

VMC is computationally cheaper than DMC and so we are able to compute
excitation energies using VMC over a larger range of system sizes.  We
expect that the finite size effects in DMC will follow those in VMC,
as our VMC calculations retrieve about 90\% of the fixed--node
correlation energy.  We have computed the $\Gamma_{25'} \! \rightarrow
\! \Gamma_{15}$ excitation energy in silicon within VMC for the system
sizes $n$ = 1, 2, 3, 4 using both promotion and addition/subtraction
methods.  The calculations were performed with ${\bf k}_s \equiv
\Gamma$ and the other computational details were the same as for the
ground state calculations.  We used Jastrow factors optimized for the
ground state of each system which were left unchanged for the excited
state.  In tests on the $n$ = 2 system we found that separately
optimizing the Jastrow factors for both ground and excited states did
not significantly change the results.  The computational cost of the
$n$ = 4 calculations is very large; an error bar in the excitation
energies of $\pm 0.3$ eV requires an error bar of $\pm 0.0006$ eV per
electron.  Although the computational effort is large we believe that
such a study is necessary to establish the accuracy of QMC excitation
energy calculations.

In Fig.~\ref{VMCexcite} we show the excitation energies obtained with
the Ewald interaction via the promotion and addition/subtraction
methods. (Results for the MPC interaction are very similar.)  The
promotion and addition/subtraction results are nearly the same, but
the promotion energies are slightly smaller because they include an
exciton binding energy, which decreases as the system size increases.
The results are consistent with a slow increase in the excitation
energy with system size and indicate that reasonable convergence is
already obtained at $n$ = 2.  The increase in excitation energies with
system size is the same trend as in HF calculations, although the
finite size errors are smaller in the correlated calculations.  The
finite size errors in the promotion and addition/subtraction methods
are not significantly different at this level of statistical accuracy.
On general grounds we expect the finite size effect in the promotion
calculation to be slightly larger.  This follows because the trend for
both promotion and addition/subtraction calculations is that the
excitation energy is reduced for small system sizes and this effect is
enhanced in the promotion calculation by the exciton binding energy
which is larger for small systems.  The calculated excitation energy
is roughly 4 eV, which is larger than the experimental value of 3.40
eV~\cite{landolt}, and also a little larger than our DMC value for an
$n$ = 2 simulation cell of 3.7 eV~\cite{bandgap}.  Our study
demonstrates that the largest contribution to the error in the VMC
band gap for $n \geq 2$ arises from the approximate nature of the
trial wave functions, and not from finite size effects.

The exciton binding energy can be calculated as the difference between
the promotion and addition/subtraction gaps.  The exciton binding
energy for the $\Gamma_{25'} \! \rightarrow \! \Gamma_{15}$ excitation
is small and we can only resolve it from the statistical noise for the
smallest ($n=1$) cell, which gives a value of $0.28 \pm 0.01$ eV.  In
earlier QMC calculations~\cite{mitas1,mitas2,mitas3} the exciton
binding energy was estimated using the Mott--Wannier formula

\begin{equation}
E_{\rm b} \sim \frac{1}{2\epsilon r}\;\;,
\end{equation}

\noindent where $\epsilon$ is the relative permittivity and $r$ is the
radius of the localisation region.  Using $\epsilon$ = 11.7 and the
appropriate radius for $n$ = 1 of $r$ = 4.0 a.u. gives an exciton
binding energy of 0.29 eV.  This is extremely close to the VMC value,
but the excellent agreement is probably fortuitous since the $n$ = 1
cell is so small that it is appropriate to use a value of $\epsilon$
at finite wave vector, which would be smaller.  The exciton binding
energy may also be evaluated within HF theory as the sum of the second
and third terms on the r.h.s. of Eq.~\ref{HF_promote_gap}.  This gives
0.75 eV for the $n=1$ cell using the Ewald interaction, which is
considerably larger than the VMC value because the latter calculation
includes screening of the electron--electron interaction.

Note that the promotion and addition/subtraction methods differ
significantly in the required computational effort. Suppose, for
example, that we wish to calculate an energy gap by either the
promotion or addition/subtraction methods.  Let us assume that the
intrinsic variance of the local energy is the same for each of the
energies, which is a good approximation for our silicon calculations,
and suppose that an acceptable error bar is obtained in a promotion
calculation by performing $M$ Monte Carlo moves for both the ground
and excited states.  A simple calculation shows that the most
efficient way to achieve the same error bar in an addition/subtraction
gap is to perform $2M$ moves for each of the $N+1$ and $N-1$ systems
and $4M$ for the ground state, giving a total cost of $8M$ moves.  It
is therefore four times more expensive to calculate an energy gap to
some given accuracy by the addition/subtraction method than by the
promotion method.

\subsection{Modified interaction for excitation energies}

In this section we introduce a modified electron--electron interaction
specifically designed to describe excitation energies within periodic
boundary conditions simulations.  Two problems arise when trying to
model excitations using finite simulation cells subject to periodic
boundary conditions.  One is that the excitation is ``squeezed'' into
the simulation cell, and the other is that there are spurious
interaction betweens the periodic replicas of the simulation cell.
Here we address the problem of the spurious interactions using the
ideas of the MPC interaction.  With our MPC interaction the replicas
interact only via the Hartree energy. The charge density on promotion
or addition/subtraction of an electron can be written as the sum of
the ground state charge density and a small deviation, i.e.,
$\tilde{\rho}({\bf r}) = \rho({\bf r}) + \Delta({\bf r})$.  We can
modify the Hartree term so that this charge density interacts with the
ground state density in the replicas.  This leads to the interaction
energy

\begin{eqnarray}\label{ultra_new}
\tilde{E}_{\rm e-e} & = & \int |\tilde{\phi}|^2\sum_{i>j} f({\bf
r}_i-{\bf r}_j) \prod_k {\rm d}{\bf r}_k \nonumber \\ & + & \int
\tilde{\rho}({\bf r})\rho({\bf r}') \left[ v_{\rm E}({\bf r}-{\bf
r}')- f({\bf r}-{\bf r}') \right] \, {\rm d}{\bf r}\, {\rm d}{\bf r}'
\nonumber \\ & - & \frac{1}{2}\int \rho({\bf r})\rho({\bf r}') \left[
v_{\rm E}({\bf r}-{\bf r}')-f({\bf r}-{\bf r}') \right] \, {\rm d}{\bf
r}\, {\rm d}{\bf r}' \;\;,
\end{eqnarray}

\noindent where $|\tilde{\phi}|^2$ generates the charge density
$\tilde{\rho}$ and the ground state charge density, $\rho$, is fixed.
A HF analysis of this interaction shows that the HF equations are
identical to Eq.~\ref{HFequation}, so that the orbitals and
eigenvalues are unaltered.  However, the ground and excited state
energy expressions are modified.  For the excited states we obtain
analogues of Eqs.~\ref{HF_promote_gap} and~\ref{koopgap}, but without
the terms involving $(v_{\rm E}-f)$, i.e., we retrieve the standard
Koopmans' theorem.  We have already shown that these terms are small
for silicon, although they will be significant in cases when the
change in the charge density due to the excitation is strongly
localized.  This analysis provides further evidence that the
electrostatic interactions between the simulation cell and its
replicas is not necessarily an important source of finite size error
in excited state energy calculations.

\section{Conclusions}
\label{sec:conclusions}
Large Coulomb finite size errors arise in total energy calculations
when using the Ewald form of the electron--electron interaction.
These finite size errors may be greatly reduced by using our MPC
interaction in which the inter--particle interaction is exactly equal
to $1/r$ at short distances and the long range interactions are
replaced by a mean--field--like one--electron potential.  It is
consistent to use the MPC interaction in conjunction with
``independent particle finite size corrections'' derived from density
functional calculations, as long as the latter calculations are
performed with the exchange--correlation functional appropriate to the
infinite system.  Although the long range mean--field--like
contribution to the MPC interaction involves the charge density of the
system, total energies are insensitive to its form and only an
approximate charge density is required.

The MPC interaction can be used consistently for all Coulomb
interactions in the system, although it normally reduces to using the
MPC interaction for the electron--electron interaction, retaining the
standard Ewald interaction for the electron--ion and ion--ion terms.
The Ewald and MPC interactions may be used in tandem as an efficient
diagnostic of Coulomb finite size errors.  If the Ewald and MPC
results agree then the Coulomb finite size error should be small.

If the simulation cell is too small then the confinement of the XC
hole makes the XC energy more negative.  This source of error is
intrinsic to using a finite simulation cell.  However, even when the
XC hole is artificially confined by a small simulation cell the MPC
interaction still gives a better estimate of the XC energy than the
Ewald interaction.

Excitation energies calculated within fixed--LDA--orbital HF theory
show significant finite size effects.  However, in correlated
calculations the finite size effects are smaller and accurate
excitation energies can be obtained using quite small simulation
cells.  In silicon we find that the finite size errors in VMC
electron--promotion (``optical absorption'') and electron
addition/subtraction (``photoemission'') calculations are similar, and
that the optical promotion method has the greater statistical
efficiency.  The finite size errors for low lying excitations in
silicon are small, and quite accurate results may be obtained from 16
atom cells.

We have described new developments aimed at understanding and reducing
finite size errors in many--body quantum simulations using periodic
boundary conditions.  Since one cannot get exact answers for an
infinite system from a finite simulation cell whatever interaction is
used, there is no ``exact interaction'' for a finite system with
periodic boundary conditions.  The Ewald and MPC interactions are
alternative model interactions compatible with periodic boundary
conditions, and the relevant question is which model interaction gives
results which most closely approximate those for very large simulation
cells.  The Ewald and MPC interactions differ by an amount which is
inversely proportional to the size of the simulation cell and
therefore they give the same energy per particle in the limit of an
infinitely large simulation cell.  However, for finite cells the Ewald
and MPC interactions can give significantly different energies.  In
every test we have performed the energy calculated with the MPC
interaction is closer than the Ewald energy to the value for a very
large system.  The MPC interaction is applicable to both metals and
insulators and it is faster to compute than the Ewald interaction.
Given these facts we believe that the MPC interaction should be used
for all quantum many--body calculations of total energies in systems
with periodic boundary conditions.

\section{Acknowledgments}

Financial support was provided by the Engineering and Physical
Sciences Research Council (UK).  Our calculations are performed on the
CRAY--T3E at the Edinburgh Parallel Computing Centre and the Hitachi
SR2201 located at the University of Cambridge High Performance
Computing Facility.

\clearpage

\begin{figure}
\caption{Convergence of the LDA and HF ground state energies per atom
of silicon as a function of simulation cell size, $n$, using
$\Gamma$--point BZ sampling.}
\label{hfldadataG}
\end{figure}

\begin{figure}
\caption{The LDA and HF ground state energies per atom of silicon as a
function of simulation cell size, $n$, using $L$--point BZ sampling.}
\label{hfldadataL}
\end{figure}

\begin{figure}
\caption{The VMC ground state energies per atom of silicon as a
function of simulation cell size, $n$.  A correction for the IPFSE is
included.  The statistical error bars are $\pm 0.01$ eV per atom.}
\label{vmcdata}
\end{figure}

\begin{figure}
\caption{The DMC ground state energies per atom of silicon as a
function of simulation cell size, $n$.  A correction for the IPFSE is
included.  The statistical error bars are $\pm 0.02$ eV per atom.}
\label{dmcdata}
\end{figure}

\begin{figure}
\caption{Convergence of the $\Gamma_{25'} \! \rightarrow \!
\Gamma_{15}$ excitation energy and valence band width of silicon.  The
data correspond to addition/subtraction energies calculated within HF
theory as a function of simulation cell size, $n$, using both the MPC
and Ewald interactions.  LDA results are also shown.}
\label{fig_koopgap}
\end{figure}

\begin{figure}
\caption{The $\Gamma_{25'} \! \rightarrow \! \Gamma_{15}$ excitation
energy of silicon calculated within VMC theory as a function of
simulation cell size, $n$.  Data for the Ewald interaction are shown
obtained via both the promotion and addition/subtraction methods.}
\label{VMCexcite}
\end{figure}


\begin{references}
\bibitem{units} All equations are given in Hartree atomic units
($\hbar=m_e=e=4\pi\epsilon_0=1$).
\bibitem{note1} The reduction of the problem to one within the
primitive unit cell is only possible within independent particle
theories; in a true many--body theory one has to solve over the entire
simulation cell.
\end{references}
\end{document}